\documentclass{article}
\usepackage{epsf,eqnfix}

\newsavebox{\LSIM}
\sbox{\LSIM}{\raisebox{-1ex}{$\ \stackrel{\textstyle<}{\sim}\ $}}

\newsavebox{\GSIM}
\sbox{\GSIM}{\raisebox{-1ex}{$\ \stackrel{\textstyle>}{\sim}\ $}}

\newcommand{\be} {\begin{equation}}
\newcommand{\ee} {\end{equation}}
\newcommand{\bdm} {\begin{displaymath}}
\newcommand{\edm} {\end{displaymath}}
\newcommand{\bc} {\begin{center}}
\newcommand{\ec} {\end{center}}
\newcommand{\beqa} {\begin{eqnarray}}
\newcommand{\eeqa} {\end{eqnarray}}

\newcommand{\bear}{\begin{eqnarray}}
\newcommand{\ear}{\end{eqnarray}}
\newcommand{\bea}{\begin{eqnarray*}}
\newcommand{\ea}{\end{eqnarray*}}

\newcommand{\slp}{\raise.15ex\hbox{$/$}\kern-.57em\hbox{$\partial$}}
\newcommand{\slG}{\raise.15ex\hbox{$/$}\kern-.57em\hbox{$G$}}
\newcommand{\slA}{\raise.15ex\hbox{$/$}\kern-.57em\hbox{$A$}}
\newcommand{\grgl}{\:\hbox to -0.2pt{\lower2.5pt\hbox{$\sim$}\hss}
{\raise3pt\hbox{$>$}}\:}
\newcommand{\klgl}{\:\hbox to -0.2pt{\lower2.5pt\hbox{$\sim$}\hss}
{\raise3pt\hbox{$<$}}\:}

\newcommand{\befi}[1]{\begin{figure}[ht] \leavevmode \centering \epsffile{#1.eps}}

\newcommand{\beq}{\begin{equation}}
\newcommand{\enq}{\end{equation}}
\newcommand{\beqast}{\begin{eqnarray*}}
\newcommand{\enqa}{\end{eqnarray}}
\newcommand{\enqast}{\end{eqnarray*}}

\newcommand{\ep}{\epsilon}

\def\surd{\sqrt}
\def\ct{\cite}
\def\pmb#1{\setbox0=\hbox{#1}
\kern.05em\copy0\kern-\wd0 \kern-.025em\raise.0433em\box0 }

\def\ell{l}

\def\P{I\!\!P}

\def\half{{\textstyle{1\over 2}}}
\def\hhalf{{1\over 2}}

\def\P{I\!\!P}
\def\pd{\partial}

\font\sevenrm=cmr7
\begin{document}

\title{Perturbative evolution at small $x$}
\author{P V Landshoff \\DAMTP, University of Cambridge}
\maketitle
\begin{abstract}
The conventional approach to perturbative evolution is illegal because
the expansion in powers of $\alpha_s$ of the the DGLAP splitting matrix 
${\bf P(z,\alpha_s)}$ breaks down at small $z$. The small-$x$ data for
the proton structure function $F_2(x,Q^2)$ and its charm component
$F^c_2(x,Q^2)$ show that a hard pomeron, with intercept close to
1.4, must be added to the familar soft pomeron. Conventional perturbative
evolution may be applied to the hard-pomeron component and provides a
striking test of perturbative QCD.
\end{abstract}

\begin{figure}[t]
\begin{center}
\epsfxsize=0.6\textwidth\epsfbox[90 580 330 775]{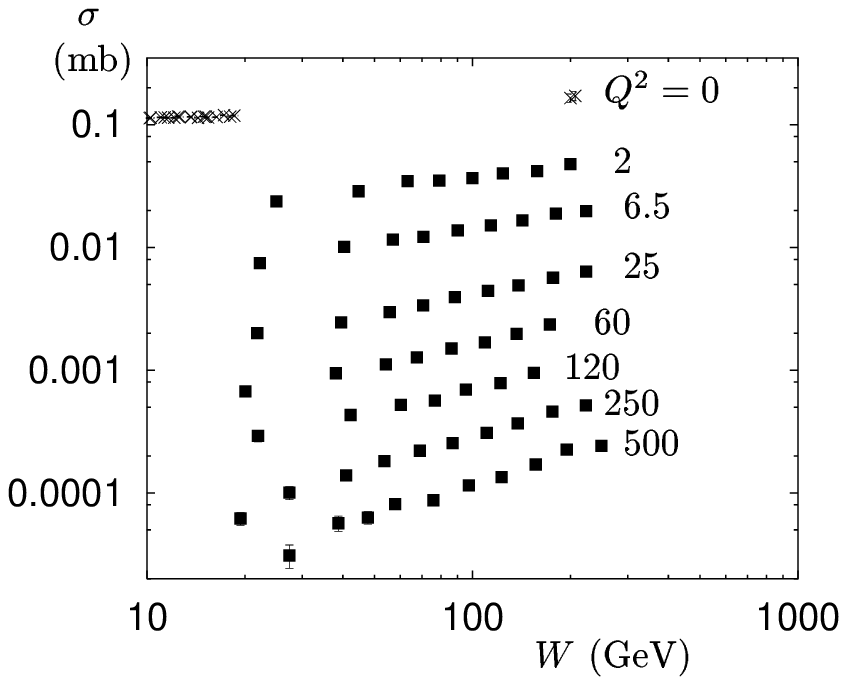}
\end{center}
\vskip -3truemm
\caption{Data\ct{H101} for $\sigma^{\gamma^* p}
=(4\pi^2\alpha_{\hbox{{\sevenrm EM}}}/Q^2)F_2$
at various values of $Q^2$,
together with real-photon data}
\label{DISSIG}
\end{figure}

\section{Introduction}

\begin{figure}[t]
\begin{center}
\epsfxsize=0.5\textwidth\epsfbox[90 580 315 760]{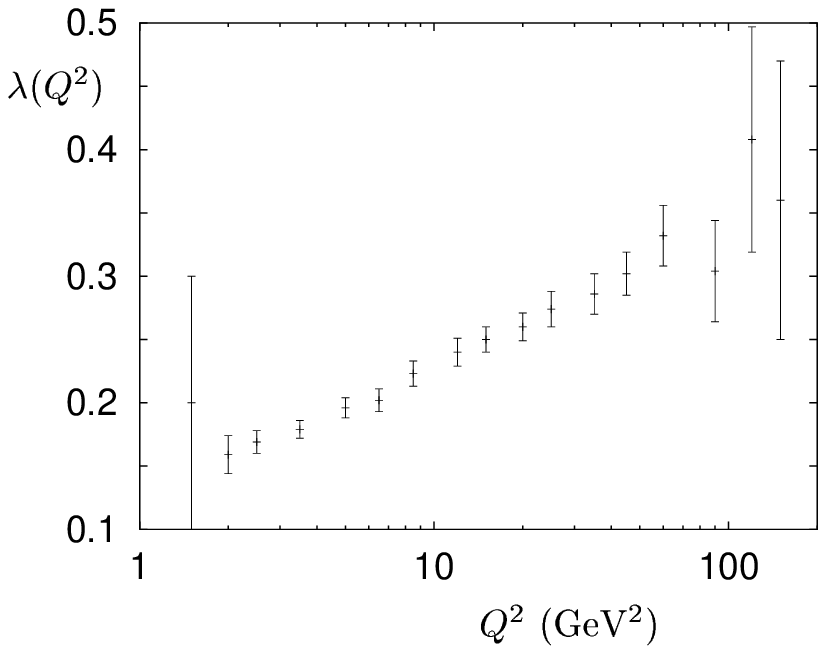}
\end{center}
\vskip -3truemm
\caption{The effective power $\lambda (Q^2)$ of(\ref{virtsig})
extracted from H1 data\ct{H1lambda}}
\label{LAMBDA}
\vskip 5truemm
\begin{center}
\epsfxsize=0.3\textwidth\epsfbox[85 670 200 770]{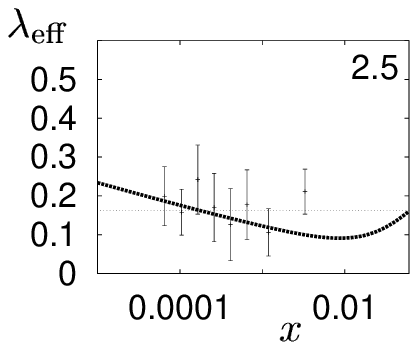}\hfill
\epsfxsize=0.3\textwidth\epsfbox[85 670 200 770]{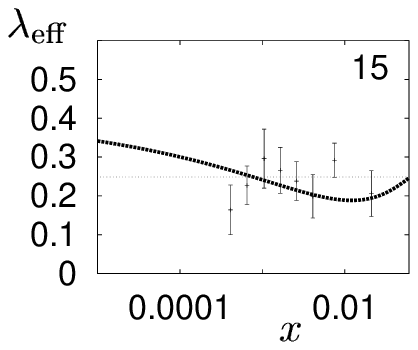}\hfill
\epsfxsize=0.3\textwidth\epsfbox[85 670 200 770]{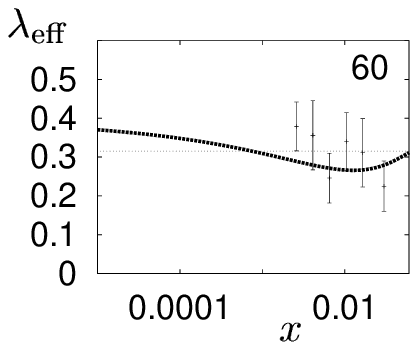}
\end{center}
\vskip -3truemm
\caption{Data\ct{H1lambda} for the effective power of $1/x$
at $Q^2=2.5,~15$ and $60$ GeV$^2$.
The horizontal lines correspond to the values plotted in figure
\ref{LAMBDA} and the solid lines to the two-pomeron fit.}
\label{POW}
\end{figure}

A striking discovery at HERA has been the rapid rise of $\sigma^{\gamma^* p}$
with $W^2$ at even quite small fixed values of $Q^2$:  
see figure~\ref{DISSIG}.
If one parametrises the rise as
an effective power
\be
\sigma(\gamma ^* p) \sim F(Q^2)\, (W^2)^{\lambda (Q^2)}
\label{virtsig}
\ee
then the power $\lambda (Q^2)$ is found to
be significantly greater than the value  0.08 that is familiar in purely
hadronic collisions\ct{DL92}.
The value of $\lambda (Q^2)$ has been extracted\ct{H1lambda} 
by the H1 collaboration
from their data and is shown in figure \ref{LAMBDA},
from which it can be seen to increase with $Q^2$ and reach about 0.4  at 
the highest values that have been measured. 

When $W^2\gg Q^2$, $x$ is small and
$W^2\sim Q^2/x$. Then the effective-power behaviour (\ref{virtsig})
corresponds to
\be
F_2(x, Q^2)\sim f(Q^2)\,x^{-\lambda(Q^2)}
\label{virtsig2}
\ee
When they extracted $\lambda(Q^2)$ from their data, to make the plot
of figure \ref{LAMBDA}, H1 assumed that the
value of $\lambda (Q^2)$ at small $x$ is independent of $x$ at each $Q^2$. 
While the data are compatible with this assumption, they do not require
it and it does not have theoretical justification. Rather, one should
parametrise the data with a sum of
fixed powers of $x$, whose relative weight varies with $Q^2$. In the
two-pomeron model I will describe, the resulting
$\lambda (Q^2)$ has significant variation 
with $x$, as is seen in figure \ref{POW}.

Even before the HERA measurements, there were predictions\cite{FR97,ESW96}
that $\lambda(Q^2)$ would be large at high values of $Q^2$. Such predictions
arose from two different equations of perturbative QCD: 
the DGLAP equation and the BFKL
equation.  However, it has since been realised that the predictions are
not as clean as initially had been hoped.

The BFKL equation is applicable when $x$ is small, and $Q^2$ 
large but not too large. 
It can be made to give values of $\lambda(Q^2)$
close to $\half$, which is approximately what experiment finds at high $Q^2$.
This led to talk of a second pomeron, the hard pomeron, with a $\lambda(Q^2)$
that might be calculated from perturbative QCD.
However, it has been realised that this calculation is almost certainly 
invalid. Apart from concerns that the approximations used to derive
the BFKL equation do not take sufficient account of energy 
conservation\cite{CL92}, it
is not correct to suppose that the equation
enables $\lambda (Q^2)$ to be calculated from
perturbative QCD alone\cite{BLV96}.
While it is conceivable\cite{BDL96,BHS97} that
there is no such problem with purely-hard processes such as
$\gamma^*\gamma^*$ collisions at very high $Q^2$, for semihard
collisions such as $\gamma^*p$ the BFKL equation inevitably 
receives important contributions from uncalculable nonperturbative effects.

The DGLAP equation 
is derived from perturbative QCD and therefore is
valid at sufficiently large $Q^2$. 
It couples the $Q^2$ variation of the quark distributions in the
proton to the gluon distribution.
When $Q^2$ is much larger than all the relevant quark masses,
the gluon distribution
affects the \hbox{$Q^2$ variation} of all the quark and antiquark
distributions equally, and the \hbox{$Q^2$ variation} of the gluon distribution
itself receives equal contributions from all the
quark and antiquark distributions. That is, the gluon distribution's
evolution is coupled to the singlet quark distribution,
which is the sum of the quark and antiquark 
distributions:
\be
q(x,Q^2)=u+\bar u+d +\bar d+c+\bar c +s+\bar s+\dots .
\label{singlet}
\ee
The number of heavy-quark terms that should be included depends on the 
value of $W$.
So the relevant form of the DGLAP equation introduces the two-component
quantity
\def\P{{\bf P}}\def\u{{\bf u}}{\def\pd{\partial}
\be
\u(x,Q^2)=\left (\matrix{q(x,Q^2)\cr g(x,Q^2)\cr}\right )\,.
\label{partonmat}
\ee
It reads
\be
Q^2{\pd\over\pd Q^2}\u(x,Q^2)=\int _x^1dz\,\P(z,\alpha_s(Q^2))\,\u(x/z,Q^2).
\label{dglap}
\ee
Here, $\P(z,\alpha_s(Q^2))$ is a $2\times 2$ matrix, called the splitting
matrix. The DGLAP equation becomes very simple if we introduce 
the Mellin transforms of $\u(x,Q^2)$ and $\P(z,\alpha_s(Q^2))$
\be
\u(N,Q^2)=\int _0^1dx\,x^{N-1}\u(x,Q^2)
\label{mellinu}
\ee
and 
\be
\P (N,\alpha_s(Q^2))=\int_0 dz\,z^N\P (z,\alpha_s(Q^2)).
\label{mellinP}
\ee
Then
\be
{\pd\over\pd t}\u (N,Q^2)= \P (N,\alpha_s(Q^2))\,\u(N,Q^2).
\label{mellin4}
\ee

In order to apply the DGLAP equation, one should first choose some 
\hbox{starting}
value $Q_0^2$ of $Q^2$ and fit the parton distributions there 
to experimental data, as functions of $x$. 
These functions cannot be calculated
from perturbative QCD, but 
the DGLAP equation, together with assumptions about
the form of the gluon distribution,
determines how they change as $Q^2$ increases. 
If\cite{GRV98} one starts at some fairly small value $Q^2_0$ of $Q^2$ 
with parton distributions that are rather flat in $x$  at small $x$, 
then an application of
the DGLAP equation results in distributions that rapidly become
steeper as $Q^2$ increases, 
in good agreement with experiment\cite{MRS98,LHK99}.  
However, this application is controversial. It is not safe to apply
a perturbative equation for choices of $Q^2_0$ as small as $1~{\rm
GeV}^2$ or less, as is sometimes done.
Indeed, it is necessary at present to
expand the splitting matrix $\P(z, \alpha_s
(Q^2) )$ in powers of $\alpha_s$ and  
this is unsafe when $x$ is small, as I explain below. 

\section{Regge approach}

A powerful approach to the small-$x$ data for $F_2(x,Q^2)$ is to extend
the Regge phenomenology that is so successful for soft hadronic
processes\ct{DL86,DL92}. Regge theory 
is not supposed to be a
competitor with perturbative QCD, but to coexist with it, and 
to be applicable when $W^2$ is much greater
than all the other variables, in particular when it is much greater than
$Q^2$. When $x$ is small $W^2\sim Q^2/x$,
so the condition that $W^2\gg Q^2$ is just that $x\ll 1$, however large
$Q^2$ may be.

Regge theory relates high-energy behaviour to singularities in
the complex angular momentum plane\ct{Col77}.
For deep inelastic scattering, the complex angular momentum $\ell$
is essentially the Mellin-transform $N$ of (\ref{mellinu}).
The correspondence is
\be
N\leftrightarrow \ell -1
\ee
so that this assumption is equivalent to assuming that the relevant 
singularities in the complex $N$-plane are simple poles.
The assumption may be not literally correct, but it turns out to give
an excellent description of the small-$x$ data\cite{DL98}. 
A pole at $N=\epsilon$ contributes a power
$x^{\epsilon}$ to the small-$x$ behaviour of $F_2$. 
So the soft pomeron contributes $x^{-\epsilon_1}$ with 
$\epsilon_1=0.08$.
But  this is not
sufficient to describe the rapid rise with $1/x$ seen in the data 
at small $x$ and large $Q^2$. Another term $x^{-\epsilon_0}$ is needed, with
$\epsilon_0\approx 0.4$. We call this $N$-plane or $\ell$-plane
singularity the hard pomeron.  This does not
explain what is its dynamical origin: maybe it is perturbative QCD though, as
I have explained, initial hopes that it might be derived
from the BFKL equation now seem unlikely to be realised.

\begin{figure}[t]
\begin{center}
\centerline{\hskip -2truemm\epsfxsize=0.484\textwidth\epsfbox[80 590 300 770]{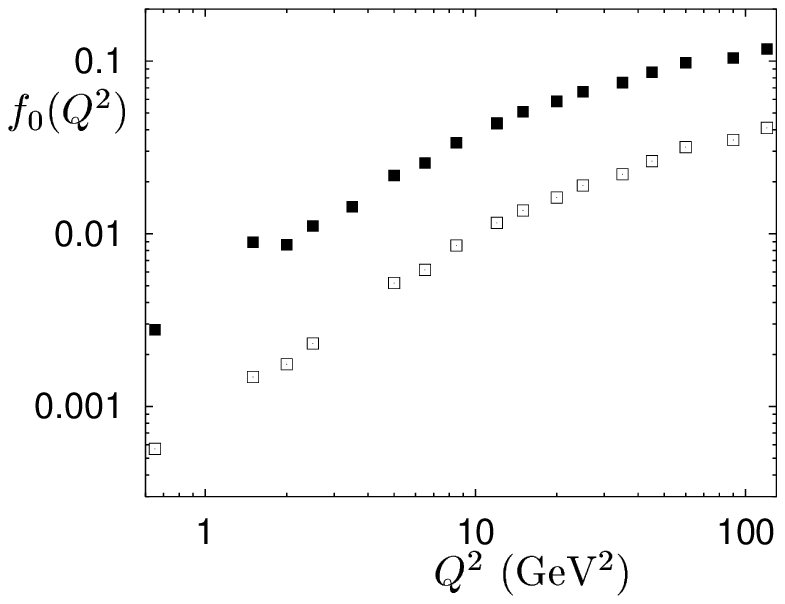}
\hfill
\epsfxsize=0.484\textwidth\epsfbox[80 590 300 770]{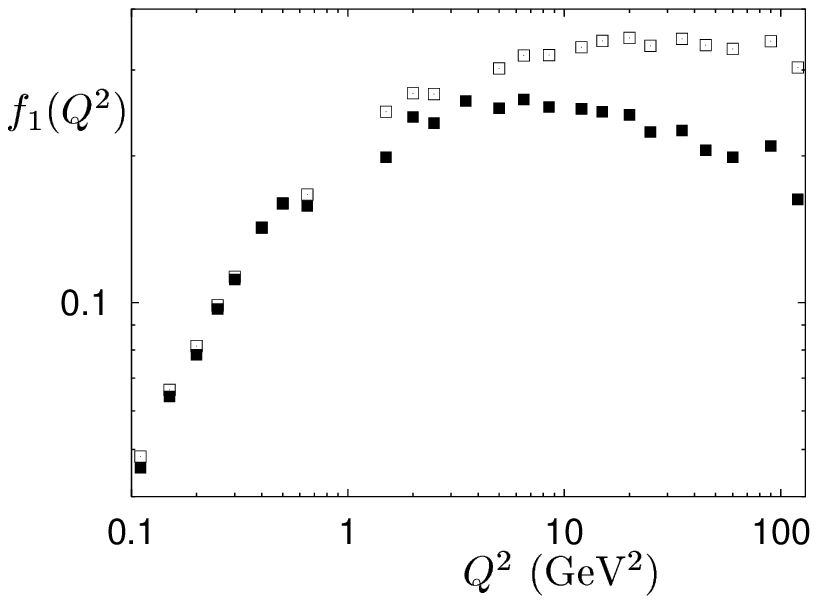}}
\end{center}
\vskip -8truemm
\caption{Fits to the coefficient functions $f_0(Q^2)$ and $f_1(Q^2)$
of (\ref{reggefit}) extracted from H1 and ZEUS data. The black points
are for $\epsilon_0=0.36$ and the white points are for $\epsilon_0=0.5$.}
\label{COEFF}
\end{figure}

So the simplest fit to the small-$x$ data corresponds to
\be
F_2 (x,Q^2 ) \sim \sum_{i=0,1} f_i (Q^2 )\, x^{-\epsilon_i}.
\label{reggefit}
\ee
where the $i=0$ term is hard-pomeron exchange and
$i=1$ is soft-pomeron exchange.
Regge theory gives no information about the form
of the coefficient functions $f_i(Q^2)$, beyond that they are
analytic functions. Also, because the real-photon cross section is
\be
\sigma^{\gamma p}={4\pi ^2\alpha\over Q^2}F_2\Bigg\arrowvert_{Q^2=0},
\label{real}
\ee
at fixed
$W$ the function $F_2$ vanishes linearly with $Q^2$ at $Q^2=0$. 
If we assume that each term in (\ref{reggefit}) has this property, then
\be
f_i(Q^2)\sim (Q^2)^{1+\epsilon _i}   ~~~~~i=0,1
\ee
near $Q^2=0$. One might hope that perturbative QCD will determine how
the two coefficient functions behave at high $Q^2$, but as I will explain
so far the theoretical
difficulties allow this only for the
hard-pomeron coefficient function $f_0(Q^2)$.
Figure \ref{COEFF} shows\ct{DL01} how $f_0$ and $f_1$ vary with $Q^2$ if we fit
(\ref{reggefit}) to the data at each $Q^2$, including values of $x$ up to 
0.02. This provides a guide on the likely functional forms of the
coefficient functions. The $f_i(Q^2)$ can then be parametrised with 
suitable functions, whose shapes resemble the data points in the plots 
and which include a number of parameters. These parameters, and the power 
$\epsilon_0$, may then be varied\ct{DL01} so as to give the best fit to all the 
small-$x$ data for $F_2(x,Q^2)$ together with the data for 
$\sigma^{\gamma p}$. This results in the value
\be
\ep_0 =0.437.
\label{epsilon0}
\ee
with
\beqa
f_0(Q^2)&=&0.0015 {(Q^2)^{1+\ep_0}\over (1+Q^2/9.11)^{1+\hhalf\ep_0}}
\label{hardcoeff}
\eeqa
and
\beqa
f_1(Q^2)&=&0.60\Big({Q^2\over 1+Q^2/0.59}\Big)^{1+\ep_1}.
\label{softcoeff}
\eeqa
The fit was made\cite{DL01} 
imposing $\ep_1=0.08$ and
using only data with $x<0.001$ and therefore $Q^2\leq 35$ GeV$^2$;
see figure \ref{F2}. With the addition of a term from $(f_2,a_2)$ exchange,
and  including also powers of $(1-x)$ in each term to make
the structure function vanish suitably as $x\to 1$, the fit agrees\ct{DL01}
remarkably well with data at larger $x$, up to $Q^2=5000$ GeV$^2$.

\begin{figure}
\begin{center}
\epsfxsize=0.48\hsize\epsfbox[90 430 290 760]{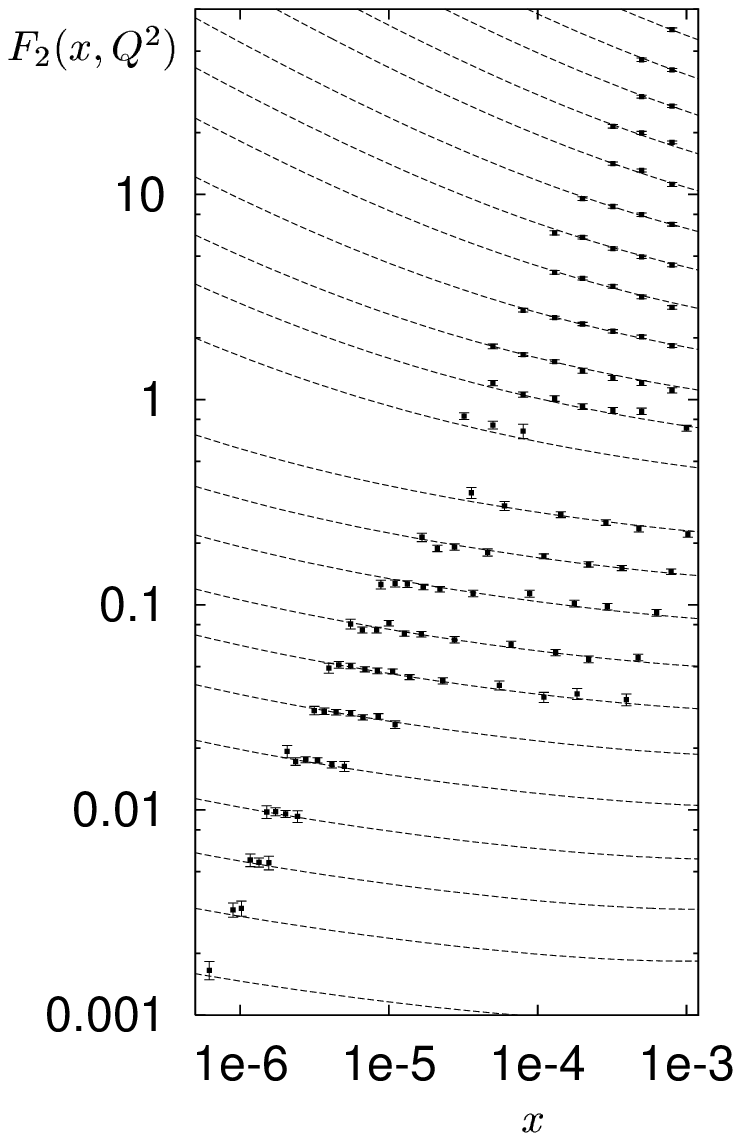}
\end{center}
\vskip -1truemm
\caption{Regge fit to data for $F_2(x,Q^2)$ for $Q^2$ between 0.045
and 35 GeV$^2$}
\label{F2}
\end{figure}

\begin{figure}[t]
\vskip -1truemm
\begin{center}
\epsfxsize=0.47\textwidth\epsfbox[85 570 330 760]{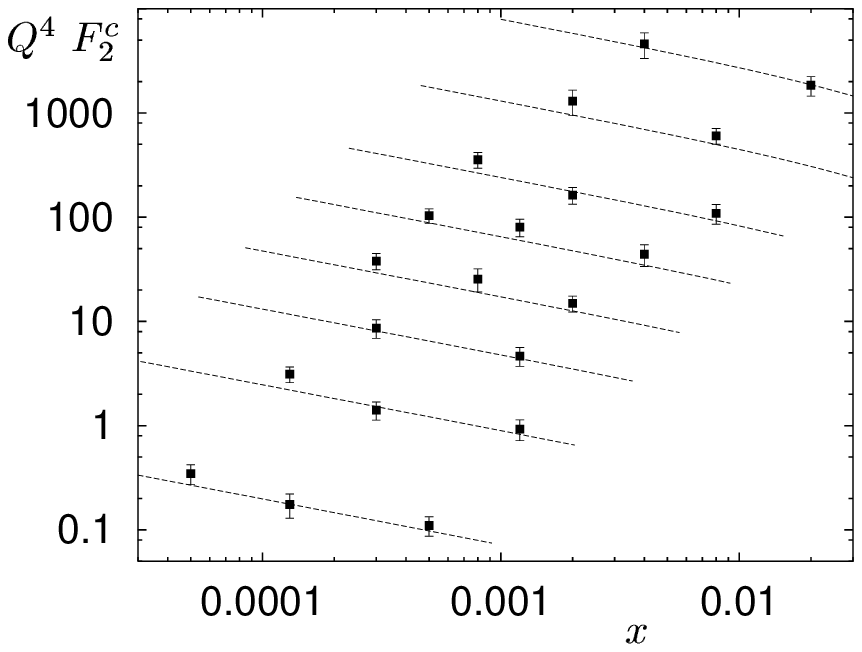}
\end{center}
\vskip -4truemm
\caption{Data for $F_2^c(x,Q^2)$ from $Q^2=1.8$ to $130$ GeV$^2$.
The lines are the hard-pomeron component of the fit to $F_2(x,Q^2)$
shown in figure \ref{F2}, normalised such that the hard pomeron is
flavour-blind.}
\label{CHARM}
\vskip 6truemm
\begin{center}
\epsfxsize=0.41\textwidth\epsfbox[50 430 290 640]{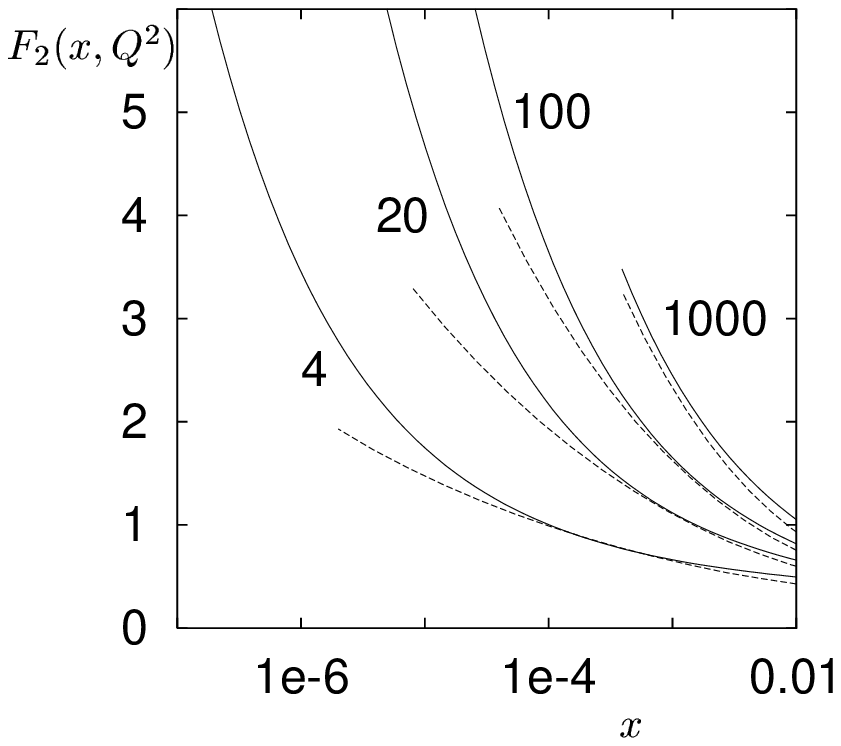}
\end{center}
\vskip -6truemm
\caption{Two-pomeron fit to $F_2(x,Q^2)$ at various values of $Q^2$
(thick curves), with two-loop unresummed perturbative-QCD fit\ct{ABF01a}
(broken curves)}
\label{EXTRAP}
\end{figure}
\begin{figure}[t]
\vskip 5truemm
\begin{center}
\epsfxsize=0.4\textwidth\epsfbox[85 430 285 760]{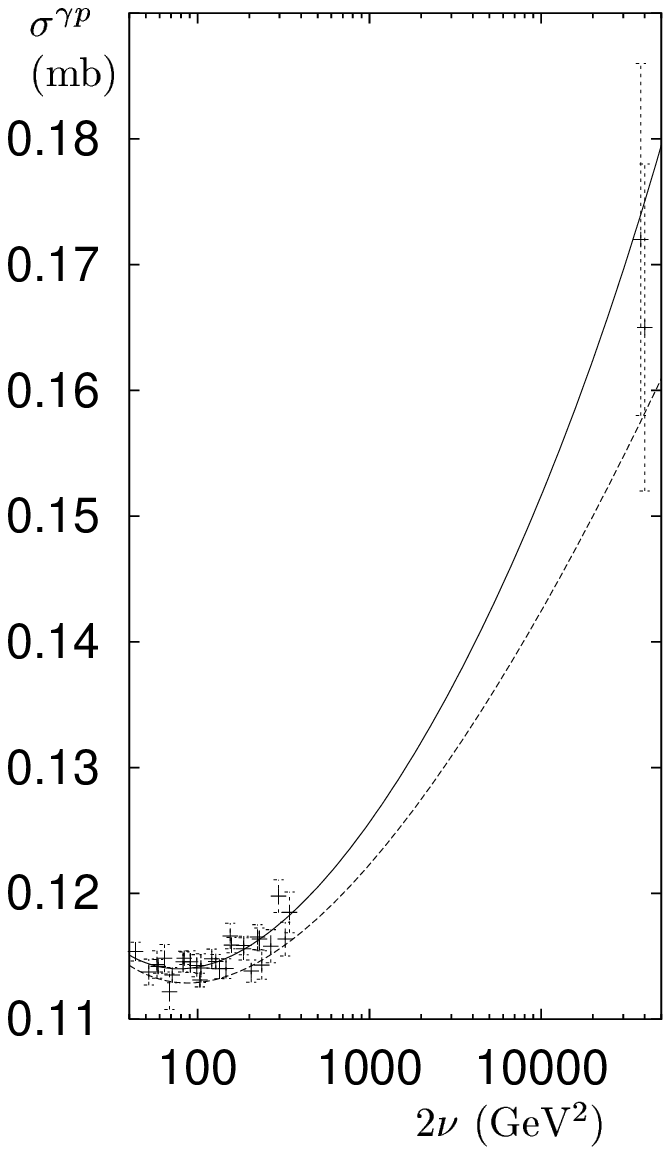}
\end{center}
\vskip -3truemm
\caption{Data for $\sigma^{\gamma p}$. The upper curve is the extrapolation
to $Q^2=0$ of the curves in figure \ref{F2}, including also
a contribution from $(f_2,a_2$ exchange; the lower curve omits
the hard-pomeron term.}
\label{REAL}
\end{figure}

Having concluded that the data for $F_2$ require a hard-pomeron component,
it is necessary to test this with other data. 
The hard pomeron is seen clearly in the charm
component of $F_2$. This describes events
in which a $D^*$ particle is produced, which are used to extract the
contribution $F^c_2 (x, Q^2 )$ to the complete 
$F_2 (x, Q^2 )$ from events where the $\gamma^*$ is absorbed by a
charmed quark.  
The data\cite{Z00} for $F_2^c (x, Q^2)$ must be treated with some
caution because the experimentalists have to make  a very large extrapolation 
to compensate for limited acceptance.  Nevertheless the data have the striking
property that, over a wide range of $Q^2$, they behave as a fixed power
of $x$:
\be
F_2^c (x, Q^2 ) = f_c (Q^2 ) x^{-\epsilon_0}
\label{charmfit}
\ee
with $\epsilon_0 \approx 0.4$: see figure \ref{CHARM}. It seems, therefore,
that $F_2^c (x, Q^2 )$ at small $x$ receives a contribution from the
hard pomeron and that the soft-pomeron contribution to
it is negligibly small. To a very good approximation the hard pomeron 
seems to be
flavour-blind\ct{DL01}: $F_2^c (x, Q^2 )$ is close to
$2\over 5$ the hard-pomeron part of $F_2(x, Q^2 )$.

For the ranges of $x$ and $Q^2$ where they overlap, the two-pomeron fit
and conventional fits based on two-loop unresummed perturbative QCD
agree with the data equally well. However, they no longer agree when they
are extrapolated to smaller values of $x$, as is seen in figure \ref{EXTRAP}.

\section{Real photons: a crucial question}

Because the data of figure~\ref{F2} include a point at extremely
low $Q^2$, it should be reliable to extrapolate these data to 
$Q^2=0$. This extrapolation is the upper curve shown in figure \ref{REAL} and, 
at $\surd s=200$ GeV, the hard-pomeron component contributes about
20 $\mu$b. That is,
at this energy, the fits with and without a hard-pomeron component differ
by about 10\%. The errors shown on the data in figure~\ref{F2} are
purely statistical; there is an additional systematic error of about
10\% at the lowest $Q^2$. So at present it is not possible to decide
whether the hard pomeron is present in the $\sigma^{\gamma p}$ data,
though it seems likely.
\begin{figure}[t]
\vskip 5truemm
\begin{center}
\epsfxsize=0.5\textwidth\epsfbox[90 590 310 765]{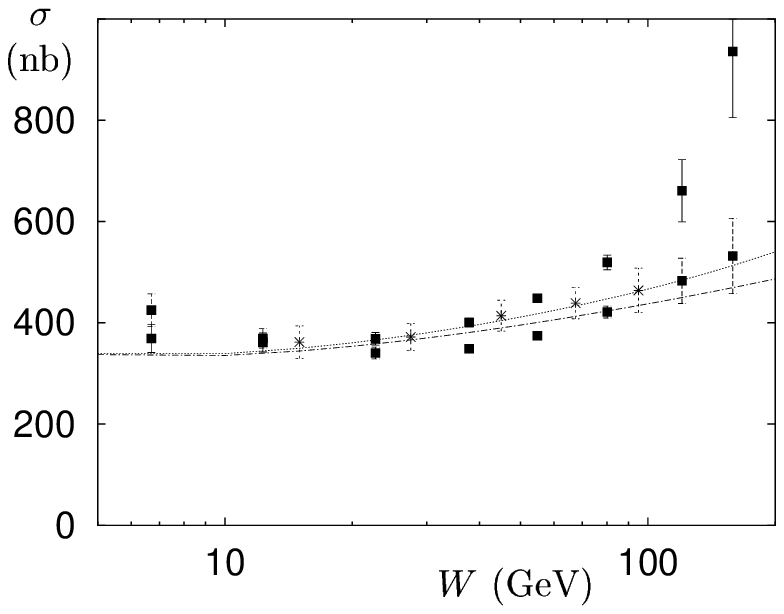}
\end{center}
\vskip -1truemm
\caption{Data for $\sigma^{\gamma\gamma}$. 
The lower curve is the 
soft-pomeron/reggeon contribution; the upper curve has an
additional hard-pomeron term.}
\label{GGHARD}
\end{figure}

\index{hard pomeron}
There is also uncertainty with the data for $\sigma^{\gamma\gamma}$
from LEP.
The cross sections are rather sensitive to the Monte Carlo model used
for the unfolding of detector effects, different Monte Carlos producing
different results. In figure \ref{GGHARD}
the resulting uncertainty is contained in the errors on
the OPAL data. The L3 data are shown with the use of two Monte Carlos, the
errors corresponding to the statistical and systematic errors combined in
quadrature.
The figure shows the prediction
obtained by applying factorisation to $\sigma^{\gamma p}$ and
$\sigma^{pp}$ data,
with no hard-pomeron component. The data from L3, particularly the
upper set, may require such a component. and the figure also shows
how adding it in might change the fit. 

This is an important question. Is the hard pomeron already present at
\hbox{$Q^2=0$}, or is it rather generated by perturbative evolution?
If it is there already at $Q^2=0$ it presumably arises because the photon
has a pointlike component, as if it is present in the $\bar pp$ total 
cross section it is very small. Maybe it will be significant at LHC
energies.

\section{Perturbative evolution}

I now show that perturbative evolution
governs how 
the hard pomeron's contribution to the structure function increases 
with $Q^2$. It is found\cite{DL02} that the parametrisation of the 
hard-pomeron coefficient function $f_0(Q^2)$ of (\ref{reggefit}) 
agrees very well with what is obtained from DGLAP evolution, over a 
large range of $Q^2$. As yet, we are not able to apply perturbative
evolution to the soft pomeron.
\begin{figure}[t]
\begin{center}
{\epsfxsize=0.58\hsize\epsfbox[75 560 350 765]{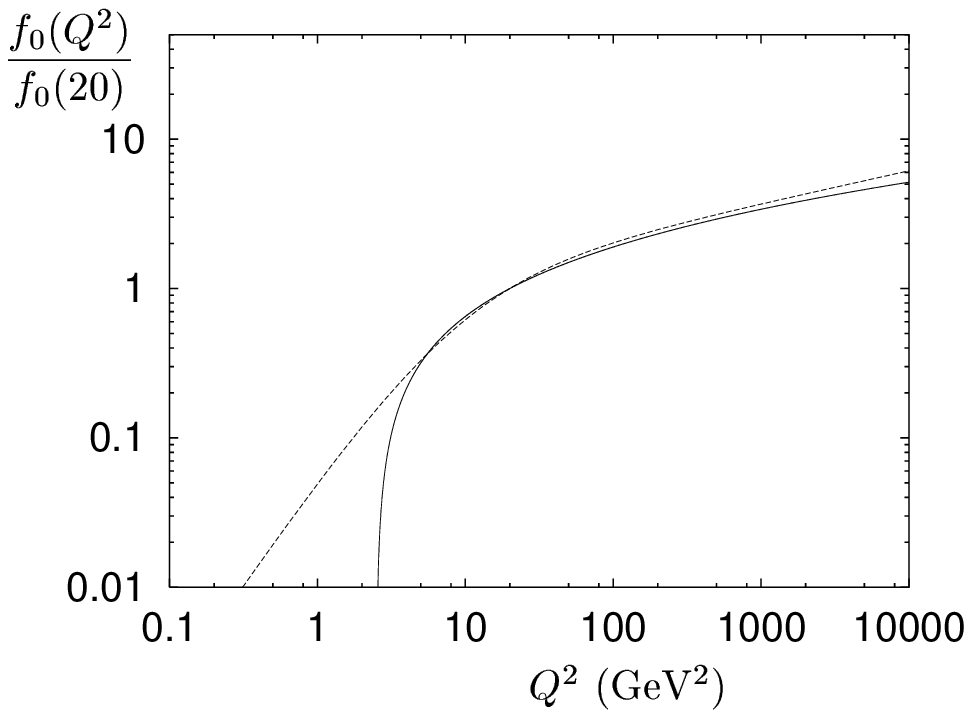}}
\vskip 3truemm
\epsfxsize=0.65\hsize\epsfbox[45 560 350 760]{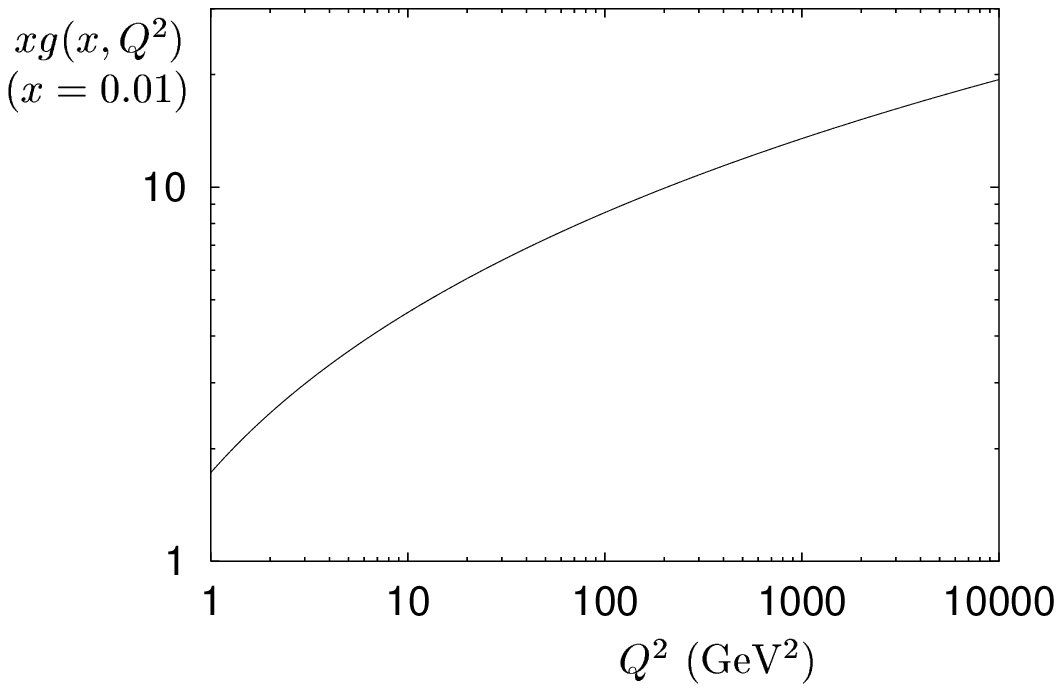}
\end{center}
\caption{(a) Next-to-leading-order
evolution with $\Lambda=400$ MeV of the hard-pomeron
coefficient $f_0(Q^2)$ (solid curve) and the fit (\ref{hardcoeff}) 
to the data (broken curve), and
evolution of the gluon structure function $xg(x,Q^2)$ at $x=0.01$.}
\label{EVOL1}
\end{figure}

\index{DGLAP equation}
\def\u{{\bf u}}\def\P{{\bf P}}\def\f{{\bf f}}
A power contribution $f(Q^2)x^{-\epsilon_0}$ to $F_2(x,Q^2)$ corresponds 
to a pole 
\be
{\f(Q^2)\over N-\epsilon_0}~~~~~~~~~~\f(Q^2)=\Big(\matrix{f_q(Q^2)\cr
f_g(Q^2)\cr}\Big )
\ee
in $\u(N,Q^2)$.  
With four active quark flavours and a flavour-blind hard pomeron,
$f_q(Q^2)={18\over 5}f_0(Q^2)$.
We find\cite{cudell}, on taking the residue of the pole at $N=\epsilon_0$
on each side of the Mellin transform (\ref{mellin4}) of the DGLAP equation,
\be
{\pd\over\pd t}\f(Q^2)= \P (N=\epsilon_0,\alpha_s(Q^2))\, \f(Q^2)
\label{fde}
\ee
If we include four flavours of quark and antiquark in the sum in 
(\ref{singlet}), then at $Q^2=20$ GeV$^2$ the singlet quark distribution
$x\sum _f(q_f+\bar q_f)\sim 0.095x^{-\epsilon_0}$ at sufficiently small
$x$.
According to figure \ref{CHARM}, the charmed-quark component $F_2^c$ of 
$F_2$ is governed almost entirely by hard-pomeron exchange 
at small $x$, even at small values of $Q^2$, and, within the 
experimental errors, its magnitude is consistent with the hard pomeron
being flavour-blind. 
According to perturbative QCD,  the charmed quark originates
from a gluon in the proton, and\ct{SN01} the two distributions are proportional
to each other to a good approximation over a wide range of $x$
and $Q^2$. This implies that the gluon
distribution also is hard-pomeron dominated. At $Q^2$=20 GeV$^2$ and
$x=0.01$, a next-to-leading-order fit\cite{H101a,C-S01} to the combined ZEUS 
and H1 data gives
$xg(x,Q^2)=5.7\pm 0.7$. Other authors\cite{MRST00,MRST01,CTEQ00}
find much the same value. This is $8\pm 1$ times the 
hard-pomeron component of the singlet quark distribution.\index{hard pomeron}
\begin{figure}
\begin{center}
{\epsfxsize=0.48\hsize\epsfbox[50 635 250 765]{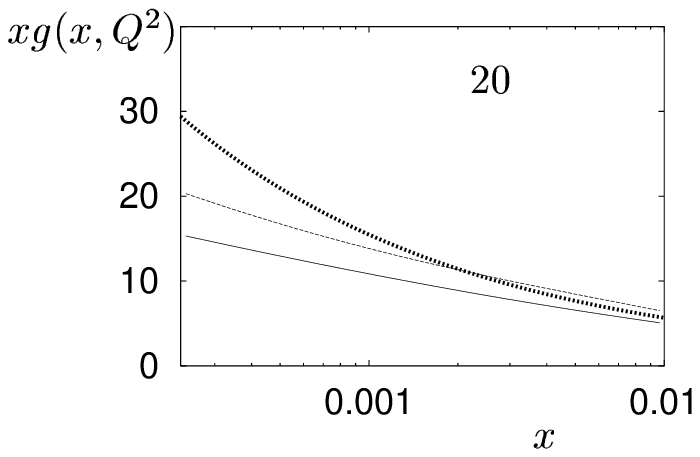}
\hfill
\epsfxsize=0.45\hsize\epsfbox[50 635 250 765]{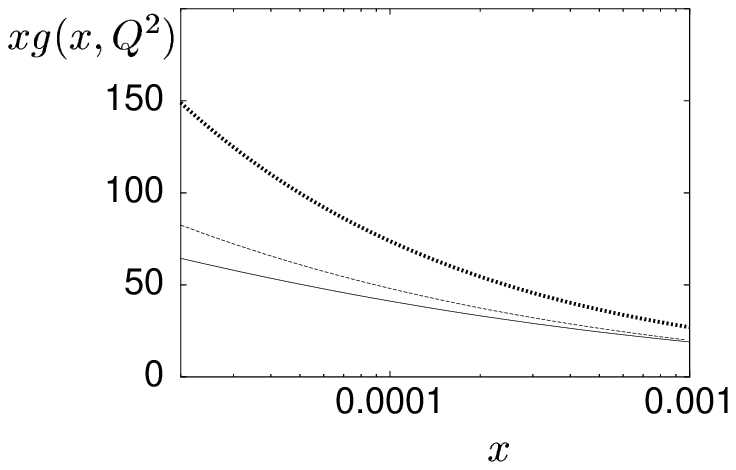}}

{\hfill (a)\hskip 70 truemm(b)$~~~~~~~~~~~~~$\hfill}
\end{center}
\vskip 2truemm
\caption{Gluon structure function $xg(x,Q^2)$ at (a) $Q^2=20$ and
(b) $200$ GeV$^2$.
In each case the thick line is our evolved distribution. In (a) the thin
lines are the limits extracted by conventional NLO analysis of HERA
data\ct{H100,C-S01}. In (b) the middle line is\ct{CTEQ00,durham} CTEQ5M and the
lower line is\ct{MRST01,durham} MRST20011.}
\label{EVOL2}
\end{figure}

An unresummed perturbation expansion 
of the splitting matrix $\P(N,\alpha_s)$ 
is not valid\cite{cudell} for small values of $N$, but we need 
$\P(N,\alpha_s)$ at $N=0.437$, that is far from 0, and so it is 
reasonable to hope that resummation is not needed.
The numerical values of the elements of the splitting matrix
$\P (N,\alpha_s)$ are known\ct{ESW96} in one and two-loop order, so it is
straightforward to integrate (\ref{fde}). At the energies being considered
it is necessary to use four flavours throughout as the charm contribution 
is active. The beauty contribution is so small that its omission has a 
negligible effect.

The result of integrating the differential equation (\ref{fde}) for the 
singlet quark distribution is shown in figure \ref{EVOL1}, where the 
solid curve is the result of the two-loop-order perturbative QCD evolution 
according to (\ref{fde}), and the broken curve is the Regge fit to the data 
I have described. The ratio of the gluon distribution to the 
hard-pomeron component of the singlet quark distribution is taken to be 
8.0 at $Q^2=20$ GeV$^2$. Figure \ref{EVOL1} also shows how the gluon distribution 
\be
xg(x,Q^2)=f_g(Q^2)x^{-\epsilon_0}
\label{gluondis}
\ee
evolves. Provided one chooses 
$\Lambda$ such that $\alpha_s(M_z^2) = 0.116$, which is the HERA 
value\cite{H101a,C-S01}, there is little difference
between the leading-order and next-to-leading-order results.

The conventional approach to evolution 
expands the splitting matrix $\P(N,\alpha_s)$ in powers of $\alpha_s$. 
Because an unresummed expansion that needs the splitting matrix
at small $N$  makes the splitting function larger than it really is, 
a gluon distribution of a given magnitude apparently gives stronger 
evolution than it really should. That is, the conventional approach 
will tend to under-estimate the magnitude of $xg(x,Q^2)$ in certain regions
of $(x,Q^2)$ space. This is verified by the results for the evolution of
$xg(x,Q^2)$ obtained from integrating (\ref{fde}). Figure \ref{EVOL2} shows 
the proton's gluon structure function at two values of $Q^2$, according to 
the solution of (\ref{fde}), which does not use the splitting matrix 
at small $N$, and compares it with what is extracted from the data by 
conventional means.

The agreement between the extraction of the hard-pomeron component
of $F_2(x,Q^2)$ from experiment, and its calculated evolution, 
is a striking success both of the hard-pomeron concept and of
perturbative QCD.
We cannot apply a similar analysis to the soft pomeron, because
this would need the splitting matrix $\P (N,\alpha_s)$ at $N=0.08$,
which is too small for an expansion in powers of $\alpha_s$
to be meaningful. We do not
yet know how to perform the necessary resummation\ct{CCS00,ABF01,Tho01}.
\vskip 10truemm
A fuller account of pomeron physics and QCD may be found in a book\ct{DDLN02}
to be published later this year by Cambridge University Press.
\vskip 7truemm
\font\eightit=cmti10 at 10truept{\eightit
This research is supported in part by the EU Programme
``Training and Mobility of Researchers", Network
``Quantum Chromodynamics and the Deep Structure of
Elementary Particles'' (contract FMRX-CT98-0194),
and by PPARC}

\bibliography{corfu}

\providecommand{\href}[2]{#2}\begingroup\raggedright\begin{thebibliography}{10}

\bibitem{DL92}
A~Donnachie and P~V Landshoff,  Physics Letters {\bf B296} (1992) 227

\bibitem{H101}
C~Adloff {\em et~al},  {\bf H1} Collaboration, European Physical Journal {\bf
  C19} (2001) 269

\bibitem{H1lambda}
C~Adloff {\em et~al},  {\bf H1} Collaboration, Physics Letters {\bf B520}
  (2001) 183

\bibitem{FR97}
J~R Forshaw and D~A Ross,  {\em Quantum Chromodynamics and the Pomeron}
\newblock Cambridge University Press (1997)

\bibitem{ESW96}
R~K Ellis, W~J Stirling and B~R Webber,  {\em QCD and Collider Physics}
\newblock Cambridge University Press (1996)

\bibitem{CL92}
J~C Collins and P~V Landshoff,  Physics Letters {\bf B276} (1992) 196

\bibitem{BLV96}
J~Bartels, H~Lotter and M~Vogt,  Physics Letters {\bf B373} (1996) 215

\bibitem{BDL96}
J~Bartels, A~{De Roeck} and H~Lotter,  Physics Letters {\bf B389} (1996) 742

\bibitem{BHS97}
J~Brodsky, F~Hautmann and D~E Soper,  Physical Review {\bf D56} (1997) 6957

\bibitem{GRV98}
M~Gl{\"u}ck, E~Reya and A~Vogt,  European Physical Journal {\bf C5} (1998) 461

\bibitem{MRS98}
A~D Martin {\em et~al},  European Physical Journal {\bf C4} (1998) 463

\bibitem{LHK99}
H~L Lai {\em et~al},  Physical Review {\bf D55} (1997) 1280

\bibitem{DL86}
A~Donnachie and P~V Landshoff,  Nuclear Physics {\bf B267} (1986) 690

\bibitem{Col77}
P~D~B Collins,  {\em An Introduction to Regge Theory}
\newblock Cambridge University Press (1977)

\bibitem{DL98}
A~Donnachie and P~V Landshoff,  Physics Letters {\bf B437} (1998) 408

\bibitem{DL01}
A~Donnachie and P~V Landshoff,  Physics Letters {\bf B518} (2001) 63

\bibitem{ABF01a}
G~Altarelli, R~D Ball and S~Forte,  hep-ph/0104246 {\bf $~$} (2001) $~$

\bibitem{Z00}
J~Breitweg {\em et~al},  {\bf ZEUS} Collaboration, European Physical Journal
  {\bf C12} (2000) 35

\bibitem{DL02}
A~Donnachie and P~V Landshoff,  hep-ph/0111427 {\bf $~$} (2001) $~$

\bibitem{cudell}
J~R Cudell, A~Donnachie and P~V Landshoff,  Physics Letters {\bf B448} (1999)
  281

\bibitem{SN01}
Z~Sullivan and P~M Nadolsky,  hep-ph/0111358 {\bf $~$} (2001) $~$

\bibitem{H101a}
C~Adloff {\em et~al},  {\bf H1} Collaboration, European Physical Journal {\bf
  C21} (2001) 33

\bibitem{C-S01}
A Cooper-Sarkar,  in {\em International Europhysics Conference on
  High Energy Physics Budapest 2001}, D~Horvath, P~Levai and A~Patkos,  eds,
JHEP (http://jhep.sissa.it/) Proceedings Section 
  PrHEP-hep2001/009 (2001) 

\bibitem{MRST00}
A~D Martin {\em et~al},  European Physics Journal {\bf C18} (2000) 117

\bibitem{MRST01}
A~D Martin {\em et~al},  hep-ph/0110215 {\bf $~$} (2001) $~$

\bibitem{CTEQ00}
H~L Lai {\em et~al},  European Physics Journal {\bf C12} (2000) 375

\bibitem{durham}
Durham data base, cpt19.dur.ac.uk/hepdata/pdf3.html

\bibitem{H100}
C~Adloff {\em et~al},  {\bf H1} Collaboration, European Physical Journal {\bf
  C21} (2001) 33

\bibitem{CCS00}
M~Ciafaloni, D~Colferai and G~P Salam,  Journal of High Energy Physics {\bf
  0007} (2000) 054

\bibitem{ABF01}
G~Altarelli, R~D Ball and S~Forte,  Nuclear Physics {\bf B599} (2001) 383

\bibitem{Tho01}
R~S Thorne,  Physical Review {\bf D64} (2001) 074005

\bibitem{DDLN02}
A~Donnachie, H~G Dosch, P~V Landshoff and O~Nachtmann,  {\em Pomeron Physics
  and QCD}
\newblock Cambridge University Press (2002)

\end{thebibliography}\endgroup
\def\bf{}
\end{document}